\documentclass[aps,prl,twocolumn,superscriptaddress,amsmath]{revtex4-1}
\usepackage{graphicx}
\usepackage{natbib}
\begin{document}
\title{Mechanical squeezing via parametric amplification and weak measurement}

\author{A. Szorkovszky}
\email[]{alexs@physics.uq.edu.au}
\affiliation{Centre for Engineered Quantum Systems, University of Queensland, Australia}
\author{A.C. Doherty}
\affiliation{Centre for Engineered Quantum Systems, University of Sydney, Australia}
\author{G.I. Harris}
\author{W.P. Bowen}
\affiliation{Centre for Engineered Quantum Systems, University of Queensland, Australia}

\date{\today}

\begin{abstract}
Nonlinear forces allow motion of a mechanical oscillator to be squeezed below the zero-point motion. Of existing methods, mechanical parametric amplification is relatively accessible, but previously thought to be limited to 3dB of squeezing in the steady state. We consider the effect of applying continuous weak measurement and feedback to this system. If the parametric drive is optimally detuned from resonance, correlations between the quadratures of motion allow unlimited steady-state squeezing. Compared to back-action evasion, we demonstrate that the measurement strength, temperature and efficiency requirements for quantum squeezing are significantly relaxed.
\end{abstract}

\maketitle

Recent experiments have demonstrated impressive progress in cooling towards the ground state and measuring the zero-point motion of mechanical oscillators. This brings within reach the observation of nonclassical phonon states, with applications in quantum information and tests of quantum mechanics\cite{blencowe1}. The most successful systems to date involve cryogenically cooled high frequency oscillators strongly coupled to optical or microwave fields\cite{teufel,anets}. However, techniques to manipulate quantum states and investigate nonclassical behaviour of phonons, apart from creating single phonon states\cite{oconnell}, are less well developed.

A squeezed state, in which the variance of one quadrature of motion is below the zero-point motion, is the most accessible of quantum resources in optomechanical systems. This can be achieved, for example, by resolved sideband cooling using squeezed or modulated input light\cite{jahne,mari}. Also promising is squeezing via back-action evading measurement (BAE)\cite{clerk}, which is close to being realized\cite{hertzberg}. These schemes would allow for ultra-sensitive force detection\cite{caves} and normal mode entanglement\cite{tian} but are constrained by the requirement of strong coupling to the optical mode. Additional downsides are the requirement of ultra-low temperatures and the side-effect of parametric instability due to strong radiation pressure\cite{hertzberg}.

Electrostatic forces, on the other hand, are strong enough to create nonclassical states in an oscillator by driving it into the nonlinear regime\cite{blencowe1}. It has been predicted that micro- and nano-electromechanical systems (MEMS/NEMS) can be engineered in this way to excite arbitrary Fock states\cite{rips} and induce macroscopic quantum tunnelling\cite{sillanpaa}. In a similar fashion, mechanical squeezing can be achieved via mechanical parametric amplification (MPA)\cite{rugar}. MPA exploits nonlinearities in the electrostatic driving field \cite{unterreithmeier}, the resonator's intrinsic motion\cite{almog} or a coupled charge qubit\cite{suh}. A periodic modulation in the oscillator's spring constant at twice its resonance frequency gives rise to an in-phase amplified quadrature and an out-of-phase damped quadrature. The amplified gain approaches infinity at threshold, whereas the squeezing due to damping is limited to a factor of one half. This is a long-standing problem that also limits the intracavity variance of an optical parametric oscillator\cite{collett}. While there are non-equilibrium schemes that allow stronger squeezing in MPA\cite{blencowe2,galve}, to our knowledge there have been no previous proposals to overcome this limit in the steady state.

\begin{figure}[!b]
\includegraphics[width=8cm]{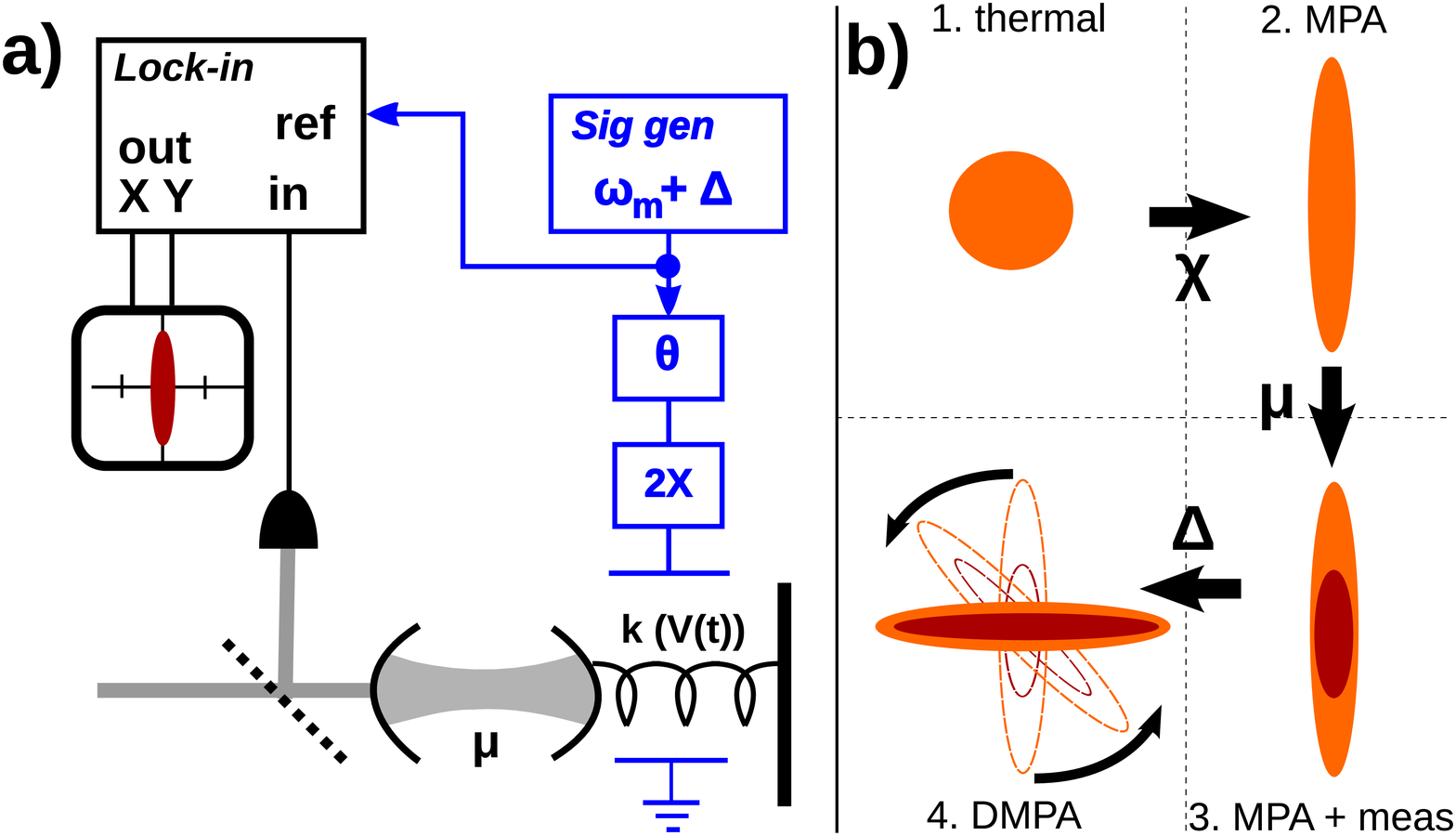}%
\caption{\label{scheme}a) Schematic of DMPA for an optomechanical system with capacitive modulation of the spring constant $k$. The measurement is mixed at $\omega_m+\Delta$ to provide the two position quadratures. b) Phase space diagrams (not to scale) depicting unequal variance of orthogonal quadratures as noise ellipses. Shown is a thermal state in a frame rotating at $\omega_m$ as a parametric drive, measurement and detuning are turned on. Darker ellipses indicate conditional variance.}
\end{figure}

Here, we consider a mechanical oscillator with detuned parametric driving and continuous quantum measurement. Such a scheme is relevant to MEMS/NEMS and optomechanical systems subject to electromechanical forces, as shown in Fig.\ \ref{scheme}a. Our approach is similar to that of \cite{woolley}, where resonant parametric driving and resolved sideband cooling are combined with quantum measurement, allowing the inference of mechanical squeezing. Our detuned MPA (or DMPA) approach is illustrated in Fig.\ \ref{scheme}b. The parametric drive amplifies the noise of one quadrature of motion, increasing the information obtained by continuous weak measurement. Detuning the drive from resonance causes the quadratures to rotate such that the measurement of the amplified quadrature contains information about the fluctuations of the squeezed quadrature, thereby further reducing its conditional uncertainty.

We find that while measurement back-action degrades the squeezing from a resonant parametric pump, a continuous weak measurement \emph{enhances} the conditional squeezing if the pump is optimally detuned. This can be made unconditional by applying appropriate negative feedback based on the measurement\cite{wiseman}. Our model predicts that current experimental parameters can produce quadrature squeezing more than 3dB below the zero-point motion without requiring an initial ground state or perfect detector efficiency. In addition, the measurement can be orders of magnitude weaker than that required for quantum squeezing via back-action evasion methods.

The system is described by the Hamiltonian for an oscillator where the spring constant $k_0=m\omega_m^2$ is modulated with amplitude $k_r$ at frequency $\omega_d = 2(\omega_m +\Delta)$,
\begin{equation}
H = \frac{\hat p^2}{2m} + \frac{\hat x^2}{2}[k_0 + k_r \cos(\omega_d t + 2\theta)]\; ,
\end{equation}
where $\omega_m$ is the mechanical resonance and $m$ is the effective mass. If the continuous measurement of position $x$ is fed into a phase-sensitive mixer such as a lock-in amplifier (see Fig.\ \ref{scheme}a), as typically done in experiments\cite{rugar}, the quadrature amplitudes $X$ and $Y$ become the variables of interest, where $\sqrt{m\omega_m/\hbar}\hat x = X\cos(\omega_m t) + Y\sin(\omega_m t)$. The phase $\theta$ between the parametric drive and lock-in detector defines the amplification axis in X-Y space. The corresponding quantum operators are $\hat X = (\hat a + \hat a^\dag)/\sqrt{2}$ and $\hat Y = -i(\hat a - \hat a^\dag)/\sqrt{2}$. In these variables, $V_X=\langle\hat X^2\rangle=V_Y=\langle\hat Y^2\rangle=1/2$ in the ground state due to the uncertainty principle. We assume low damping $\gamma\ll \omega_m$ corresponding to a high mechanical quality, and small perturbation of the spring constant $k_r\ll k_0$, so that a rotating wave approximation can be made in the interaction picture at the frequency $\omega_d$, giving
\begin{equation}\label{ham}
\tilde H = -\hbar\Delta \hat a^\dag\hat a + i\hbar \frac{\chi}{2}(e^{2i\theta}\hat a^2-e^{-2i\theta}\hat a^{\dag2}) \; .
\end{equation}
The second term in this Hamiltonian takes the form of a squeezing operator proportional to $\chi=\omega_m k_r/2k_0$. The nonlinearity $\chi$ can be interpreted as the mechanical frequency shift due to a change in spring constant by $k_r$.

The conditional evolution of the system, including back-action and thermal noise, is described by a stochastic master equation for continuous position measurement\cite{jacobs}. If the measurement rate $\mu$ is small compared to $\omega_m$, the rotating wave approximation can be performed on this master equation using the method of Ref.\ \cite{wiseman2}. The expectation value of an arbitrary observable $\hat A$ is then found to evolve as
\begin{eqnarray}\label{master}
\mathrm{d}\langle\hat A\rangle & = & -\frac{i}{\hbar}\langle[\hat A,\tilde H]\rangle\,\mathrm{d}t + [2\gamma N + \mu]\langle\mathcal{D}[\hat a^\dag]\hat A\rangle\,\mathrm{d}t\nonumber \\
& & + [2\gamma(N+1)+\mu]\langle\mathcal{D}[\hat a]\hat A\rangle\,\mathrm{d}t \\ 
&  & + \sqrt{\eta \mu}\langle\mathcal{H}[\hat X]\hat A\rangle\,\mathrm{d}W_X + \sqrt{\eta\mu}\langle\mathcal{H}[\hat Y]\hat A\rangle\,\mathrm{d}W_Y \; , \nonumber
\end{eqnarray}
where $N$ is the mean thermal phonon number, $\eta$ is the quantum efficiency and $\mathrm{d}W_X$ and $\mathrm{d}W_Y$ are uncorrelated Wiener processes corresponding to the two position quadratures\cite{wiseman2}. It is assumed that the measurement signal also has no thermal fluctuations, which is valid for optical readout or a well-cooled signal amplifier. The superoperators $\mathcal{D}$ and $\mathcal{H}$ are defined as
\begin{eqnarray}
\mathcal{D}[\hat a]\hat A & = & \hat a^\dag\hat A\hat a - \frac{1}{2}(\hat a^\dag \hat a\hat A + \hat A\hat a^\dag \hat a) \nonumber \\
\mathcal{H}[\hat a]\hat A & = & \hat a\hat A + \hat A\hat a^\dag - \langle \hat a+\hat a^\dag\rangle\hat A \; . \nonumber
\end{eqnarray}
Setting $\theta=\Delta=0$ and substituting $\hat X$ and $\hat Y$ for $\hat A$ in Eq.\ (\ref{master}) results in damping rates of $\gamma+\chi$ and $\gamma-\chi$ respectively for the two quadratures. When there is no nonlinearity ($\chi=0$), these reduce to the bare damping rate $\gamma$ as expected. There exists a threshold for self-oscillation at $\chi=\gamma$, where the damping of $\hat Y$ is zero, corresponding to infinite gain. At this threshold, the damping of $\hat X$ is double its original value, resulting in a gain of $1/2$ and therefore a noise squeezing limit of 3dB.

Non-zero detuning $\Delta$ of the pump (and lock-in reference) from resonance rotates the phase of the oscillations, creating elliptical trajectories of the mean values and resulting in a shift in the steady-state squeezing axis\cite{carmichael}. Most importantly, as we will show, correlations between the quadratures allow the squeezed quadrature to be more efficiently localized by measurement. The modified below-threshold condition is $\chi^2 < \Delta^2 + \gamma^2$ where above this, the detuning and damping are insufficient to keep the mean values decaying to zero, causing instability. With sufficient detuning, this theory is limited only by the rotating wave approximation, which requires the condition $\chi \ll \omega_m$.

Near threshold, the squeezing is maximized at an angle approaching $-\pi/4$ \cite{carmichael}. Therefore, for simplicity we set the drive phase $\theta=\pi/4$ so that the squeezing axis approximately corresponds to $X$. The evolution of the variances $V_X$,$V_Y$ and covariance $C$ can be derived from the master equation using It\=o calculus and the relations
\begin{gather}
\mathrm{d}V_A = \mathrm{d}\langle \hat A^2\rangle - 2\langle \hat A\rangle\mathrm{d}\langle \hat A\rangle - (\mathrm{d}\langle \hat A\rangle)^2 \nonumber \\
\mathrm{d}C = \frac{1}{2}\mathrm{d}\langle \hat X\hat Y\!+\!\hat Y\hat X\rangle \!-\! \langle \hat X\rangle\mathrm{d}\langle \hat Y\rangle \!-\! \langle \hat Y\rangle\mathrm{d}\langle \hat X\rangle \!-\! \mathrm{d}\langle \hat X\rangle\mathrm{d}\langle \hat Y\rangle \nonumber
\end{gather}
All terms involving $\mathcal{D}$ in Eq.\ (\ref{master}), including thermal noise and a back-action term $\mu$, add to these position variances. The superoperator $\mathcal{H}$, however, results in the conditional variance decreasing as the effective measurement rate $\eta\mu$ increases. Setting the differentials above to zero yields the steady state solutions as three coupled equations
\begin{gather}
\label{var1} V_X = \frac{\sqrt{\gamma^2 + z - 8\eta\mu C(\Delta - \chi) - (4\eta\mu C)^2} - \gamma}{4\eta\mu} \\
\label{var2} V_Y = \frac{\sqrt{\gamma^2 + z + 8\eta\mu C(\Delta + \chi) - (4\eta\mu C)^2} - \gamma}{4\eta\mu} \\
\label{var3} C  = \frac{\chi(V_Y+V_X)-\Delta(V_Y-V_X)}{4\eta\mu(V_Y+V_X)+2\gamma} \; .
\end{gather}
Here, the conditioning parameter $z=8\eta\mu\gamma(N+N_{BA}+1/2)$ characterises how well the motion is detected and $N_{BA}=\mu/2\gamma$ is the additional phonon number due to back-action for continuous measurement. When the pump is off, the covariance vanishes and both quadratures have the standard back-action limited variance conditioned on the measurement result
\begin{equation}\label{meas}
V_0 = \frac{\sqrt{\gamma^2 + z} - \gamma}{4\eta\mu}\; , \\
\end{equation}
which in the weak measurement limit ($\mu\ll\gamma$) gives the expected thermal state variance $V_T = N + 1/2$. The conditional variance decreases with stronger measurement and saturates at the quantum limit of $1/2$ in the limit $\mu\gg\gamma$, where the ground state motion can be resolved.

The parametric pump creates a non-zero covariance, or squeezing at an angle of $\pi/4$ to the $X$ and $Y$ quadratures, while the detuning rotates the squeezing axis so that $V_X<V_Y$. For given parameters, there exists a detuning $\Delta_{opt}$ that optimally uses the measurement to minimise $V_X$. When the measurement is switched off ($z=0$), this is exactly solvable as $\Delta_{opt}-\chi = \gamma$ with a variance limit of $V_0/2$, agreeing with time domain analysis of intracavity parametric squeezing in optics\cite{collett}.

A more general analytic result can be obtained for the optimal squeezing in the limit that the conditioning parameter $z\gg \gamma^2$, corresponding to high temperature and/or strong measurement. Solving for the minimum squeezed variance $V_X$ over all detunings results in the solution
\begin{equation}\label{varsol}
V_{Xopt} \approx \frac{V_0}{2}\sqrt{2 + 3(\Delta'_{opt}-\chi')^2 - \frac{1}{(\Delta'_{opt}-\chi')^2}} \; ,
\end{equation}
where dashed parameters are normalized by $\sqrt{\gamma^2+z}$ and the optimal detuning $\Delta_{opt}$ is given by
\begin{equation}\label{deltasol}
\Delta_{opt}'-\chi' = \frac{1}{6}(\mathrm{Re}\{G\} + \sqrt{3}\mathrm{Im}\{G\}-3\chi') \; ,
\end{equation}
with
\begin{equation}\label{deltasol2}
G = \left(27\chi'^3 + 6\sqrt{3}i\sqrt{27\chi'^4+36\chi'^2+16}\right)^{\frac{1}{3}} \; .
\end{equation}
\begin{figure}
\includegraphics[width=8cm]{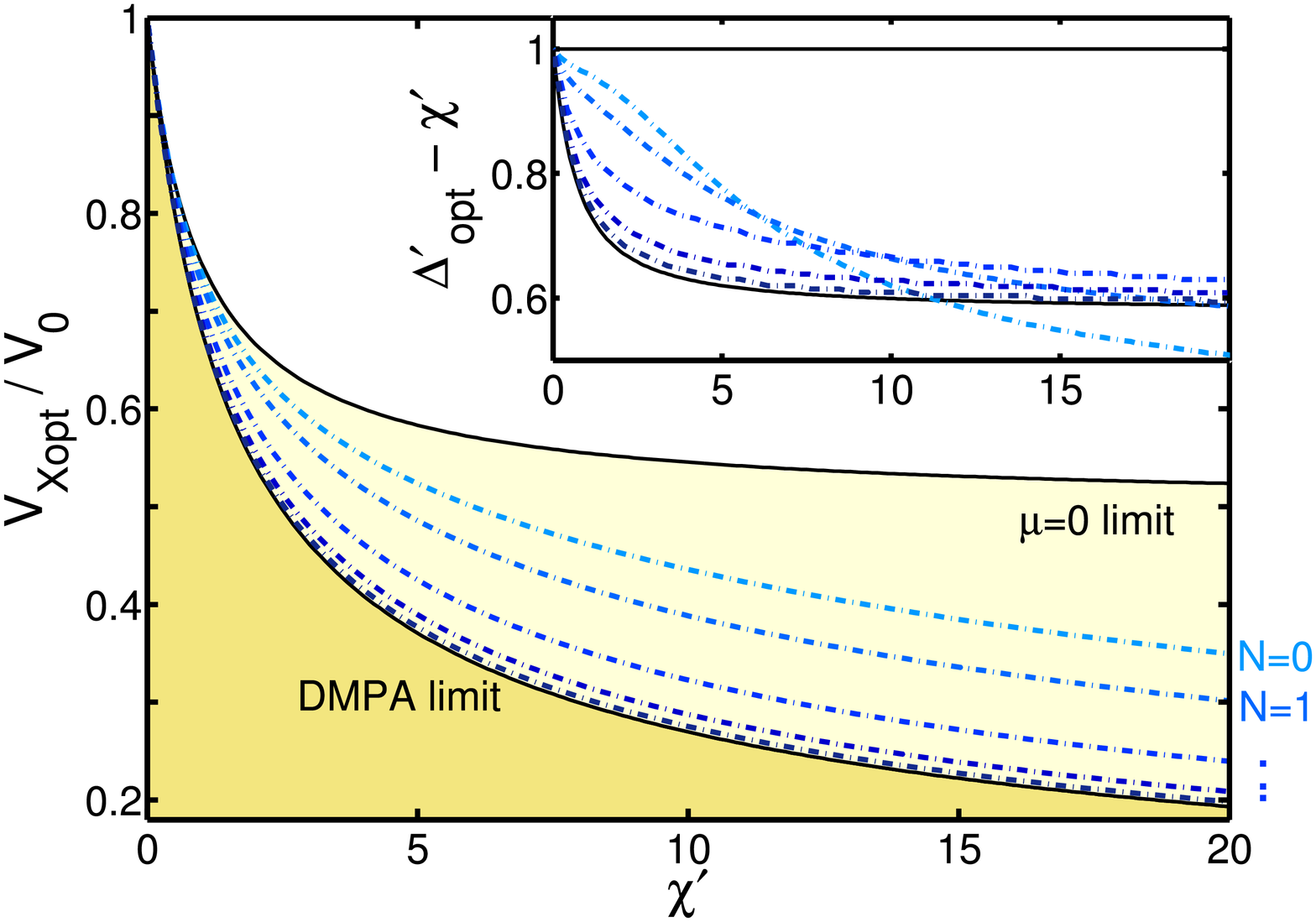}%
\caption{\label{hightemp}Optimal $X$ quadrature squeezing and (inset) optimal detuning as a function of nonlinearity $\chi'$. Dotted lines: numerical solutions for measurement strength $\mu/\gamma = 0.1$ and mean phonon number $N=0,1,10,100,1000$. Solid lines: analytic solutions for no measurement (upper) and for the limit $z\gg\gamma^2$ (lower).}
\end{figure}
It can be seen from Eqs (\ref{varsol}-\ref{deltasol2}) that $\Delta_{opt}'-\chi'$ approaches 1 in the weak pump limit $\chi\ll\gamma$, giving $V_{Xopt}=V_0$ as expected. In the strong pump limit, $\Delta_{opt}'-\chi'\rightarrow1/\sqrt{3}$ and hence $V_X/V_0\rightarrow 0$. This shows that with quantum measurement, improvement on the conditional variance is not bounded by $1/2$ and perfect squeezing is possible with sufficient nonlinearity $\chi'$. Fig.\ \ref{hightemp} compares the analytic solutions for $V_{Xopt}$ and $\Delta_{opt}$ to numerical solutions for a fixed measurement strength of $\mu/\gamma=0.1$ and various initial phonon numbers $N$. The analytic solutions fit well at high temperatures and provide a lower bound to the squeezing.

The normalized nonlinearity $\chi'$ is a useful figure of merit for classical squeezing at high temperature. In the limit where back-action is negligible this can be expressed, in terms of mechanical properties, as
\begin{equation}
\chi' \approx \frac{Q}{4\sqrt{2N\eta\mu/\gamma}}\frac{k_r}{k_0}  \; ,
\end{equation}
where $Q$ is the mechanical quality factor $\omega_m/\gamma$. Hence for mechanical squeezing, a high Q and low spring constant $k_0$ is desirable. If $N\eta\mu/\gamma$ is large (i.e.\ the thermal noise is well transduced) the statistics are dominated by the measurement conditioning and the effect of the parametric drive is reduced. However, lower temperature always results in lower absolute variance $V_{Xopt}$ as expected.

\begin{figure}
\includegraphics[width=8cm]{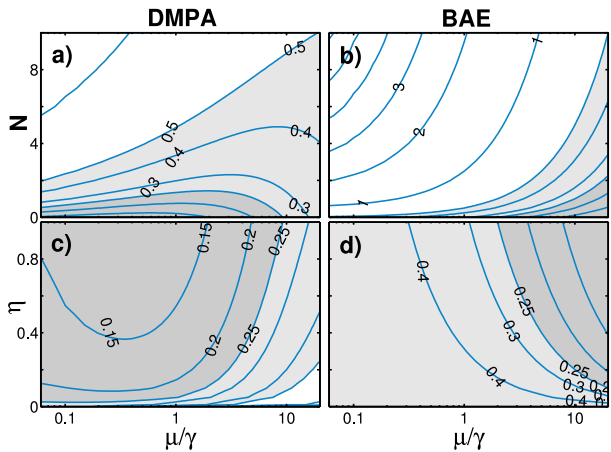}%
\caption{\label{gndstate} Optimal quantum squeezing generated by DMPA with $\chi/\gamma=50$ (a,c) and by BAE (b,d). $V_{Xopt}$ is shown as a function of $\mu/\gamma$ and $N$ in (a,b) for $\eta=1$, and as a function of $\mu/\gamma$ and $\eta$ in (c,d) for $N=0$. Shaded areas denote squeezing below the zero-point motion (light) and the standard 3dB MPA limit (dark).}
\end{figure}

For low initial temperatures and weak measurement, $z<\gamma^2$ so the above solutions are no longer applicable. Therefore, in order to investigate \emph{quantum} squeezing ($V_X<1/2$), Eqs (\ref{var1}-\ref{var3}) were solved numerically for initial temperatures near the ground state. The absolute variance $V_X$ is shown in Fig.\ \ref{gndstate} for $\chi/\gamma=50$, compared to that achievable by BAE (given by Eq.\ (\ref{meas}) where $N_{BA}=0$ \cite{clerk}). For relatively weak measurements, the DMPA method produces variances well below the zero-point motion. This method is also robust to heating, with squeezing possible from $N<5$ at $\mu/\gamma\approx1$. At this measurement strength, a BAE scheme would require $N<0.5$. Furthermore, the need for an often impractical modulation of the measurement strength is eliminated.

It should be noted that in the limit of very low temperatures and strong measurement, where back-action degrades the DMPA squeezing, BAE is preferable. The turning point occurs at a measurement strength of $\mu/\gamma \approx 0.4$ for $N=0$. At higher temperatures, stronger measurement is required to reach the back-action dominated regime. For $N=0$ and $\chi/\gamma=50$, 6dB of squeezing is achievable using DMPA, twice that achievable without detuning and measurement. This is robust against detection inefficiency, with a reduction to 40\% efficiency only degrading the squeezing to 5dB. Unconditional squeezing can be achieved by using a linear feedback force (e.g.\ from a separate electrode) to stabilize the mean values\cite{wiseman}. This relies on high mechanical Q so that the delay in converting momentum feedback into physical displacement can be neglected.

We have shown that a detuned parametric amplifier with weak measurement can vastly reduce the uncertainty of one quadrature of motion of a mechanical oscillator. At low initial temperatures, this manifests as quantum squeezing. This method therefore opens up the possibility of generating mechanical squeezed states in many systems, namely those that can be cooled to low phonon occupations and parametrically driven independently of the measurement apparatus.

High frequency mechanical oscillators approaching the GHz regime have the well-known advantage of being able to be cooled to very low phonon numbers\cite{oconnell}. In addition, the small scale involved facilitates a lower intrinsic spring constant $k_0$, especially with extreme dimensional ratios such those found in oscillators based on carbon nanotubes\cite{huttel}. Many current NEMS resonators are already capable of above-threshold parametric amplification and are at high enough frequency to be cooled near the ground state. For example, the 9MHz oscillator used in Ref.\ \cite{unterreithmeier} can reach $\chi/\gamma=50$ with a drive amplitude of $0.6$V. Other suitable examples could involve using electrostatic forces to drive NEMS resonators coupled to optical cavities\cite{anets,rips}, or even to drive the internal mechanical modes of optical cavities \cite{lee,sridaran}. These setups allow resolved sideband cooling of the mechanical mode\cite{schliesser} as well as the necessary parametric drive.

\begin{acknowledgments}
This research was funded by the Australian Research Council Centre of Excellence CE110001013 and Discovery Project DP0987146.
\end{acknowledgments}

\end{document}